\documentclass[useAMS,usenatbib]{mn2e}
\usepackage{psfig}
\newcommand{\uchii}{{UCH{\sc\,ii}}}

\title[The sensitivity of line profiles]
{The Sensitivity of Infall Molecular Line Profiles to the Ambient
Radiation Field}
\author[M.P. Redman et al.]
{M.P. Redman$^{1,2}$, J.M.C. Rawlings$^{1}$, J.A. Yates$^{1}$ and D.A. Williams$^{1}$ \\
$1$Department~of~Physics~and~Astronomy, University~College~London,
Gower Street, London WC1E 6BT, UK\\
$1$School of Cosmic Physics, Dublin Institute for Advanced Studies, 5
Merrion Square, Dublin 2, Ireland\\}

\date{Received **insert**; in original form **insert**}
\pagerange{\pageref{firstpage}--\pageref{lastpage}}
\begin{document}
\label{firstpage}
\maketitle

\begin{abstract} 
In cold molecular clouds submillimetre emission lines are excited by
the ambient radiation field. The pumping is dominated by the cosmic
microwave background (CMB). It is usual in molecular line radiative
transfer modelling to simply assume that this is the only incident
radiation field.  In this paper, a molecular line transport code and a
dust radiative transfer code are used to explore the effects of the
inclusion of a full interstellar radiation field (ISRF) on a simple
test molecular cloud. It is found that in many galactic situations,
the shape and strength of the line profiles that result are robust to
variations in the ISRF and thus that in most cases, it is safe to
adopt the CMB radiation field for the molecular line transport
calculations. However, we show that in two examples, the inclusion of
a plausible radiation field can have a significant effect on the the
line profiles. Firstly, in the vicinity of an embedded massive star,
there will be an enhanced far infared component to the
radiation field. Secondly, for molecular clouds at large redshift, the
CMB temperature increases and this of course also alters the radiation
field. In both of these cases, the line profiles are weakened
significantly compared to a cloud exposed to a standard radiation
field. Therefore this effect should be accounted for when investigating
prestellar cores in massive star forming regions and when searching
for molecular clouds at high redshift.
\end{abstract}


\section{Introduction} 
Molecular line profiles from prestellar and protostellar objects can
potentially yield dynamical information about the collapse process
that leads to the formation of stars. In the last decade or so, the
common practice has been to obtain line profiles from an infall
candidate and then to use a radiative transfer code, together with a
dynamical model to interpret the data. Inferences are then made about the
validity of one or other of the competing star formation models.
However, it is becoming clear that the interpretation of these line profiles
can be fraught with difficulty. \citet{rawlings&yates01} used a
self-consistent chemical and dynamical model of collapsing
star-forming cores to explore the effects of abundance
variations. They showed that the line profiles are very sensitive to
the assumed values of the free parameters in the chemical models. 

\citet{ward-thompson&buckley01} have presented a
sensitivity analysis of the line profiles to various free parameters,
in the context of modelling the Class 0 sources, NGC1333-IRAS2 and
Serpens SMM4. The {\sc stenholm} radiative transfer code (see,
e.\@g.\@~\citealt{heaton.et.al93}) was used, adopting analytical fits
to the \citet{shu77} collapse model for the velocity and density
profiles, and an optically thin thermal balance model for the dust
temperature.  The following free parameters were investigated: the
impact parameter, or beam offset from the central position, the beam
size (or assumed source distance), the infall velocity, the fractional
abundances of the tracer species (assumed not to vary with position),
the turbulent velocity and the systemic rotational velocity. One of
the most important (general) findings was that the qualitative shape
of the line profiles - in particular the velocity separation between
the red- and blue-shifted peaks in the self-reversed line profiles is
highly sensitive to the turbulent velocity.  This results from the
broadening of the absorption profiles of the foreground envelope,
whilst the total integrated flux increases with the turbulent velocity
dispersion as the effective optical depth of the emitting gas is
reduced. 

In this paper, the results of numerical experiments to investigate the
sensitivity of molecular line profiles to the radiation field incident
on a realistic test globule are reported. Firstly, the effects on
molecular line profiles of changing this field from the CMB to a more
realistic ISRF is investigated. Secondly, the ISRF is modified to
simulate the radiation field in the vicinity of embedded massive
stars. Finally the ISRF is modified by increasing the CMB temperature,
to simulate the radiation field in which molecular clouds at
high-redshift will be found. Our study lacks the more self-consistent
approach of the combined hydrodynamical/chemical/radiative transfer
models of Rawlings and Yates (2001), but serves to show the
sensitivities and hence to highlight the diagnostic strengths (and
weaknesses) of line profile analyses.

The discussion is limited to two well known molecular tracers used to
diagnose collapsing cores: HCO$^+$ (J=3$\to$2) and $^{13}$CO
(J=2$\to$1, 3$\to2$). This allows direct comparison with other work;
previous studies (eg. Rawlings and Yates, 2001) have shown that the
HCO$^+$ transition is particularly likely to be strongly self-reversed
and asymmetric, whilst the effects on the CO lines has implications
for high-redshift molecular observations. The discussion is also
limited to zero beam offset positions so as to allow a simple
comparison of the effects discussed in this paper.

\section{Radiative transfer modelling of submillimetre line emission}
The radiative transfer code used in this work is {\sc smmol}, an
approximate $\Lambda-$iterative code that solves multi-level non-LTE
radiative transfer problems (see \citealt{rawlings&yates01} for full
details).  The results generated by this code can be viewed with some
confidence because a series of recent benchmarking exercises have
compared several such codes, including for example {\sc ratran}, an
accelerated Monte-Carlo code described by
\citet{hogerheijde&vandertak00}. Both {\sc smmol} and {\sc ratran},
despite the very different numerical methods used, are able to yield
the same results when applied to test cases very similar to the model
runs described here. The benchmarking has been described by
\citet{vanzadelhoff.et.al02}.

For our numerical studies, a test globule was constructed using the
class 0 source B335, which is one of the best observed (and modelled)
protostellar infall candidates
(e.\@g.\@~\citealt{zhou.et.al93,choi.et.al95,wilner.et.al00}). The
physical parameters of the test globule (following Rawlings \& Yates
2001) are based upon the models of B335 of Zhou et al. (1993) and Choi
et al. (1995). At each of fifty points in a radial grid (similar to
that of \citealt{choi.et.al95}) the radius, density, fractional
abundance, gas temperature, dust temperature, radial velocity and
microturbulent velocity are specified. The density and velocity
profiles [$n(r)~\mbox{and}~v(r)$] are taken from the Shu (1977)
inside-out collapse model, with the physical parameters as specified
in Table 1. B335 is being used because there is a well developed model
for it, but it is not claimed that B335 itself is being subjected to
any of the effects explored in this paper.

Modifying the radiation field incident on the globule will affect the
dust temperature. To account for this, the dust temperature in the
globule is calculated for each radiation field using the dust
radiative transfer code described in
\cite{efstathiou&rowan-robinson94}. For the case of a standard ISRF at
z=0, it was verified that the radial profile of the dust temperature
is comparable to that deduced from dust continuum observations of B335
(Zhou et al., 1990). As in Rawlings and Yates (2001), we assume that
the dust and gas are thermally well coupled. Although this assumption
is probably not valid in the low density ($n<10^5~{\rm cm}^{-3}$)
outer envelope, we again note that it holds in the denser line-forming
core regions.  In fact, the dust/gas temperature changes turn out to
only have a second order effect on the line profiles compared with
changing the radiation field. The collapse model is not
well-justified, and more reliable evaluations of the temperature
profile are now available
\citep{shirley.et.al00} but this approach provides a
useful and well-known benchmark against which to test the effects
discussed in this paper.

Most of the transitions that we consider in this study are very
optically thick, so that the line profiles are relatively insensitive
to the absolute values of the fractional abundances ($X_{\rm
i}$). This result is in marked difference to the findings of
Ward-Thompson and Buckley (2000) who found strong qualitative and
quantitative differences when $X_{\rm i}$ was varied between
$10^{-9}-10^{-8}$. We suspect that this is a result of the inability
of the {\sc stenholm} code to deal with extreme optical depths.
However, as emphasised in Rawlings and Yates (2000), radial {\it
variations} in the abundances can have profound effects on the line
profiles. We do not want to confuse the interpretation of our results
with chemical issues, so we simply adopt a two-phase chemical scheme,
with separate abundances within and exterior to the infall radius
($r_{\rm inf}$); Choi et al. (1995) used a single constant abundance
in their work. The abundances used in our model - which are spatial
averages taken from fig. 1 of Rawlings and Yates (2000), are also
given in Table 1.

{\small
\begin{table*}
\begin{tabular}{lll}
\hline
Cloud radius & 0.15 pc & \\
Collapse age ($t_{\rm coll}$) & $1.36\times10^5$ years & \\
Infall radius ($r_{\rm inf}$) & 0.03 pc & \\
Distance to source & 250 pc & \\
Telescope diameter & 15 m & \\
Beam efficiency & 0.6 & \\
$X$(HCO$^+$) & $5\times 10^{-9}$ ($r < r_{\rm inf}$) &  
$5\times 10^{-10}$ ($r > r_{\rm inf}$) \\ 
$X$($^{13}$CO) & $1\times 10^{-7}$ ($r <r_{\rm inf}$) & 
$2\times 10^{-7}$ ($r >r_{\rm inf}$) \\ 
\hline
\end{tabular}
\caption{Parameters used in the dynamical and radiative transfer model.}
\end{table*}} 

\section{Inclusion of a full interstellar radiation field}
Most molecular line transport codes adopt the CMB as the incident
radiation field \citet{vanzadelhoff.et.al02}. While this is the
dominant source of radiation at the wavelengths of the molecular
transitions under consideration here, a proper treatment should
include the full interstellar radiation field. \citet{vandishoeck94}
has reviewed how such a radiation field is constructed. It contains
three dominant components: starlight (mainly from B stars),
dust-enshrouded massive stars, and the CMB. It is these latter two
components that are modified in this work and discussed below. Firstly
though, the effect of replacing the CMB with a realistic ISRF was
investigated. In this work, the radiation field constructed by
\citet{evans.et.al01} was employed. This is a combination of the
radiation field introduced by \citet{black94} with that of
\citet{draine78}. The dust radiative transfer code was used to firstly
calculate the gas/dust temperature before the line profiles were
generated by the molecular line transport code.

The inclusion of the Black-Draine ISRF had almost no discernible
effect on the line profile shapes for both species. This is because
the CMB component of the radiation field vastly dominates the line
formation at these wavelengths. Therefore, in most galactic
environments, it is safe to simply adopt the CMB as the only
significant incident radiation field. The ISRF of
\cite{mathis.et.al83} is a plausible alternative radiation field but
using this produced little discernible difference in the
line profiles. However, the dust continuum strength will be affected by the
UV and IR photons and is sensitive to the exact form of the radiation
field used.

\section{Varying the incident radiation field on a galactic globule}
The embedded massive star component in the ISRF is particularly
important since the few forming massive stars are typically
accompanied by huge numbers of low mass star forming objects. The
radiation peak from an embedded massive star typically occurs at
around $100~{\rm \mu m}$ and can be approximated as a blackbody of
temperature $\sim 30~{\rm K}$ (e.\@.g.\@~\citealt{chini.et.al87}). In
diffuse clouds the contribution of the massive star radiation field is
several orders of magnitude less than the CMB at the wavelengths of
molecular line tracer transitions, simply because of the small
covering fraction on the sky of these regions. However, in the
vicinity of a hot core or embedded \uchii\ region the fraction of the
sky occupied by the cloud may be large enough that the radiation field
at mm wavelengths is enhanced.

\begin{figure}
\psfig{file=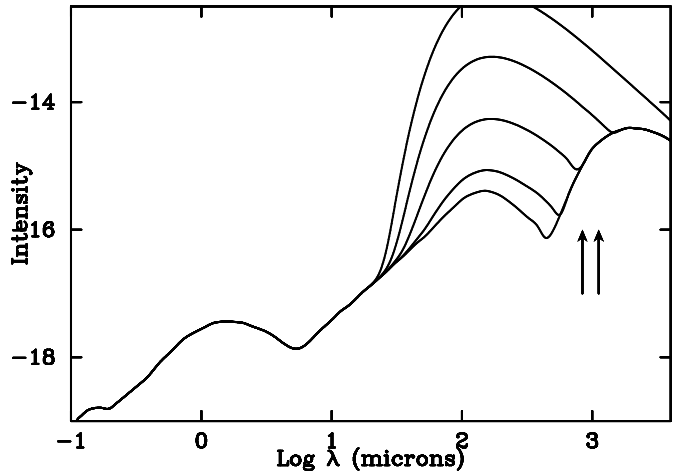,angle=270,width=250pt,bbllx=332pt,bblly=391pt,bburx=587pt,bbury=747pt}
\caption{Black and Draine radiation field with an additional component due to a neaby embedded massive star. This is represented by a 30 K blackbody, scaled by factors of $10^{-4}$, $10^{-3}$, $10^{-2}$, $10^{-1}$ to represent variations in the fractional sky coverage of the embedded object. The two arrows mark the
wavelengths of the ${\rm HCO^+}~J=3\rightarrow 2$ and $J=4\rightarrow
3$ lines. The intensity scale is logarithmic with units of ${\rm erg~s^{-1}~cm^{-2}~Hz^{-1}~sr^{-1}}$}
\label{30varblackdraine}
\end{figure}
Figure~\ref{30varblackdraine} is a plot of the Black-Draine ISRF to
which has been added a 30 K grey body to represent the increased flux
due to an embedded massive star component. The ratio of the grey body
to black body flux is varied between $10^{-2}$ and $3\times
10^{-1}$. Of course, using an isotropic grey body flux is a crude
simplification of the real situation of an adjacent localised
radiation source, a point returned to in Section 6.  The two arrows
mark the wavelengths of the ${\rm HCO^+}~J=3\rightarrow 2$ and
$J=4\rightarrow 3$ lines.  It can be seen that once the flux ratio
exceeds around $10^{-2}$, the mm continuum flux at the wavelengths of
the lines is altered significantly.

\begin{figure}
\psfig{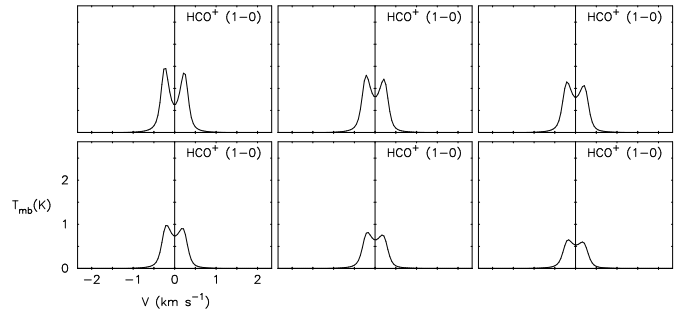}
\caption{Model HCO$^+~1\rightarrow 0$ line profiles for a Black-Draine radiation field
with a diluted 30 K blackbody. From top left to right the dilution
factor has the following values: $0.01, 0.1, 0.15, 0.2, 0.25, 0.3$}
\label{uchiifirst}
\end{figure}
\begin{figure}
\psfig{file=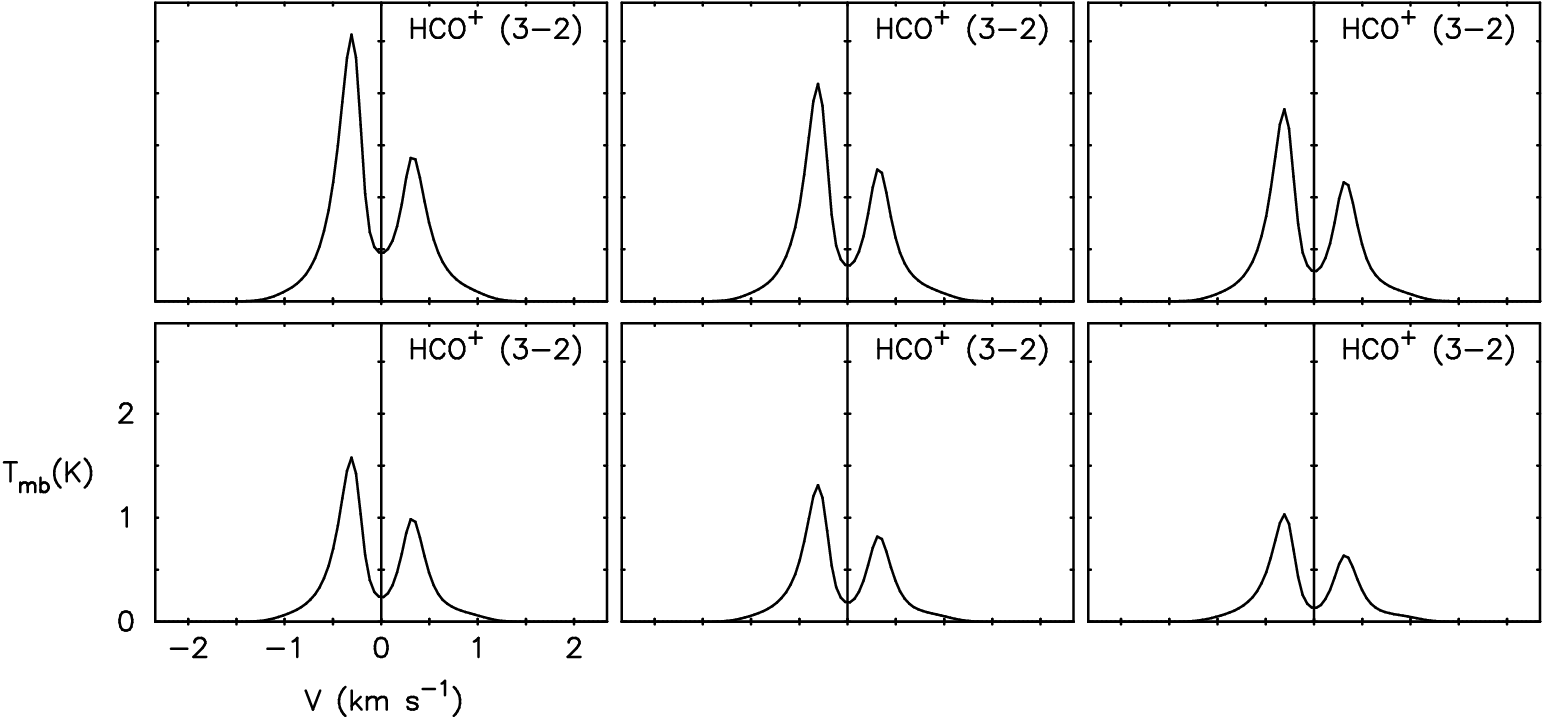,angle=270,width=250pt,bbllx=66pt,bblly=21pt,bburx=411pt,bbury=762pt}
\caption{Model HCO$^+~3\rightarrow 2$ line profiles for a Black-Draine radiation field
with a diluted 30 K blackbody. From top left to right the dilution
factor has the following values: $0.01, 0.1, 0.15, 0.2, 0.25, 0.3$}
\label{uchii2}
\end{figure}
\begin{figure}
\psfig{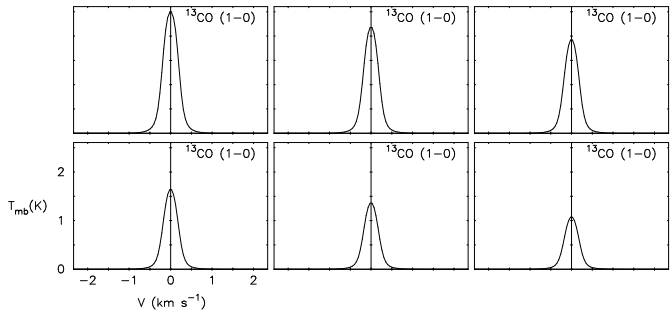}
\caption{Model $^{13}{\rm CO}~3\rightarrow 2$ line profiles for a Black-Draine radiation field
with a diluted 30 K blackbody. From top left to right the dilution
factor has the following values: 0.01, 0.1, 0.15, 0.2, 0.25, 0.3}
\label{uchii3}
\end{figure}
\begin{figure}
\psfig{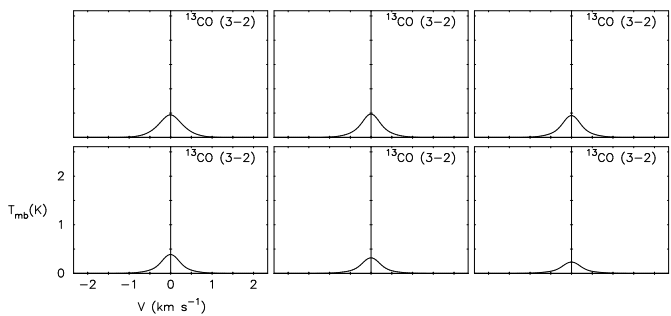}
\caption{Model $^{13}{\rm CO}~1\rightarrow 0$ line profiles for a Black-Draine radiation field
with a diluted 30 K blackbody. From top left to right the dilution
factor has the following values: 0.01, 0.1, 0.15, 0.2, 0.25, 0.3}
\label{uchiilast}
\end{figure}

Figures~\ref{uchiifirst}-\ref{uchiilast} are $^{13}$CO and HCO$^+$
line profiles from the test globule for the different radiation
fields.  Inspection of the profiles indicates that the line strengths
are significantly reduced when the globule is exposed to increasing
ambient radiation. The effect is slightly more pronounced in the
lowest transitions of both of the species. Similar results obtain for
calculations of the line profiles of CS, which for brevity are not
presented here. In other words, if a cloud is close enough to an
embedded massive star that over around one percent of the sky is
occupied then it will not be valid to simply adopt a CMB radiation
field: the diluted embedded massive star component must be added to it
first. For example, the sizes of the envelopes surrounding massive
stars are of order $\sim 10^{18}~{\rm cm}$ and so any starless cloud
within a few parsecs will be exposed to a significant flux of mm
photons from the tail of the blackbody from the embedded star. Thus
any future studies of individuals from the thousands of low mass stars
that form in the vicinity of high mass stars should take this effect
into account.

\section{Varying the incident radiation field on a high redshift globule}
\begin{figure}
\psfig{file=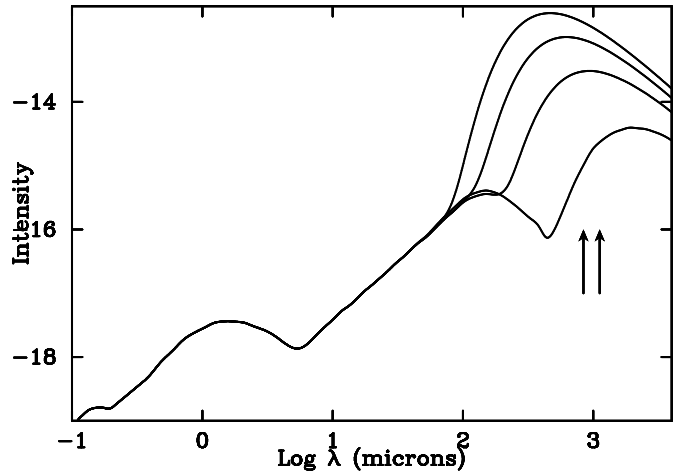,angle=270,width=250pt,bbllx=332pt,bblly=391pt,bburx=587pt,bbury=747pt}
\caption{Black and Draine radiation field at high redshift. The CMB component increases with redshift. Displayed are the radiation fields for z=0 (lowermost curve), 1, 2, and 3 (uppermost curve). The two arrows mark the wavelengths of the ${\rm HCO^+}~J=3\rightarrow 2$ and $J=4\rightarrow
3$ lines. The intensity scale is logarithmic with units of ${\rm erg~s^{-1}~cm^{-2}~Hz^{-1}~sr^{-1}}$.}
\label{cmbrad}
\end{figure}

Star formation appears to have peaked at a redshift of around $z=1$
and appears to have been occuring as early as $z=6.4$
(Walter~et~al.\@~2003, Bertoldi~et~al.\@~2003
\nocite{bertoldi.et.al03,walter.et.al03} and see e.\@g.\@
\citealt{illingworth99} for a review of the subject). The first
appearance and evolution of molecular gas in galaxies over this time
is also very uncertain. \citet{norman&spaans97} have discussed the
formation of molecules at high redshift. While at present the
observational difficulties of studying high redshift molecular clouds
are severe, it is nonetheless interesting to consider the effects of
the warmer CMB radiation field at earlier epochs and its effect on any
molecular clouds present then. To investigate this, the Black and
Draine radiation field above was modified by varying the CMB
temperature component with redshift such that
\begin{equation}
\frac{T_z}{T_{\rm cmb}} = 1 + z,
\end{equation}
where $T_{\rm cmb}$ is the current CMB temperature of 2.728 K. Values
of $z$ up to 3 were considered. The shapes of the radiation field
produced by these values of $z$ are displayed in Figure~\ref{cmbrad}. As
before, the modification of the internal temperature structure by dust
heating produces only second order changes to the line profile
shape. The most significant changes are caused by the change in CMB
temperature, as discussed below.

With the model specified as above, we considered variations in the
incident FIR and mm-wave radiation field on the cloud as the only free
parameter. This is initially provided by the Cosmic Microwave
Background (CMB) radiation, characterised by a black-body of
temperature 2.728K. The radiation is assumed to be isotropic and
illuminates the exterior of the cloud. We alter the value of the FIR
and mm-wave radiation field by specifying different radiation
temperatures ($T_{\rm rad}$) corresponding to redshifts of 0, 1, 2,
and 3 according to equation 1.

Varying the incident radiation field in this way has an even more
dramatic effect than the changes described in Section 4 - the line
profiles of each of the three tracers become progressively weaker as
the temperature rises, until at the highest temperatures considered in
this study ($\simeq 11~\rm K$) they almost vanish. Higher temperature runs
(not presented here) were also carried out and show that the profiles
even move into absorption with increasing temperature.

\begin{figure}
\psfig{file=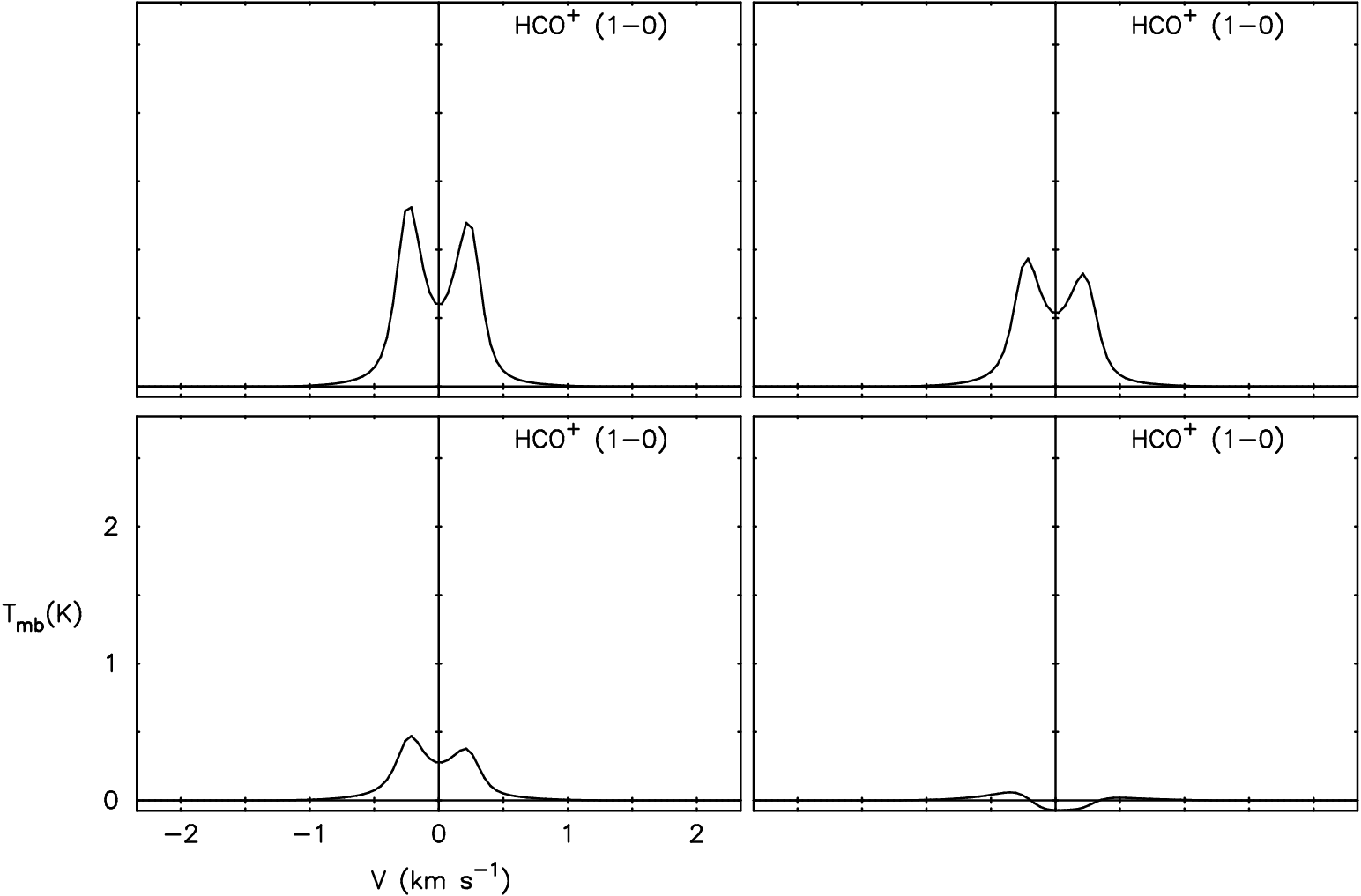,angle=270,width=250pt,bbllx=66pt,bblly=21pt,bburx=554pt,bbury=762pt}
\caption{
Model HCO$^+~1\rightarrow 0$ line profiles for a Black-Draine
radiation field modified such that the CMB temperature is that
appropriate for high redshift. From top left to right the redshifts are: 0, 1, 2, 3}
\label{cmbfirst}
\end{figure}

\begin{figure}
\psfig{file=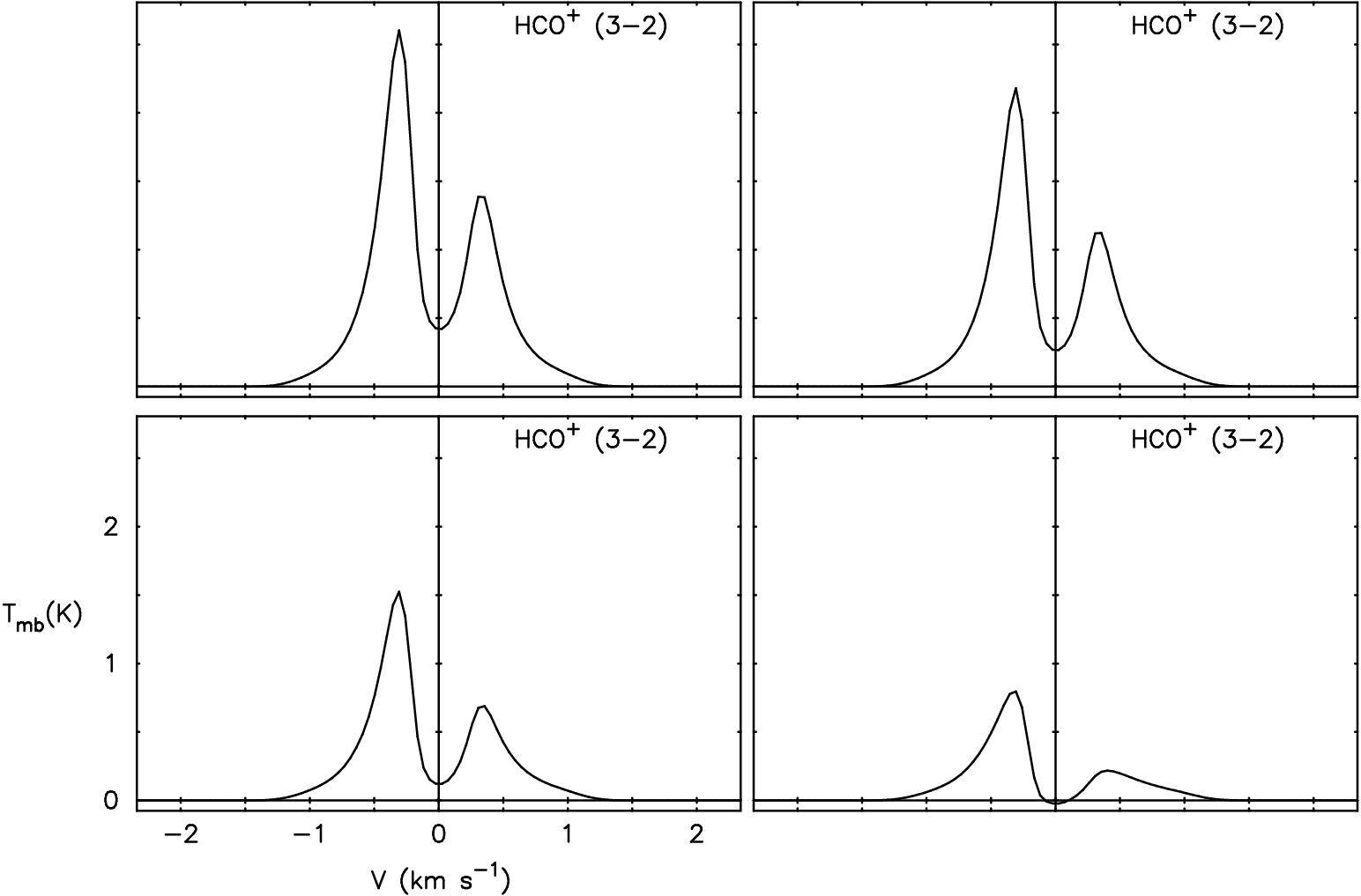,angle=270,width=250pt,bbllx=66pt,bblly=21pt,bburx=554pt,bbury=762pt}
\caption{
Model HCO$^+~3\rightarrow 2$ line profiles for a Black-Draine
radiation field modified such that the CMB temperature is that
appropriate for high redshift. From top left to right the redshifts are: 0, 1, 2, 3}
\label{cmb2}
\end{figure}

Figs~\ref{cmbfirst}-\ref{cmblast} show the continuum-subtracted line
profiles generated for the HCO$^+$ (J$=3\to2$) and (J$=1\to0$) and the
$^{13}{\rm CO}$ (J$=3\to2$) and (J$=1\to0$) transitions respectively,
for different values of the external radiation temperature. The
gradual suppression of the emission line profile with increasing
temperature is immediately clear. The numerical experiment shows how a
bright FIR/mm-wave radiation field can greatly diminish $^{13}$CO
emission.

\begin{figure}
\psfig{file=MB687f9.eps,angle=270,width=250pt,bbllx=66pt,bblly=21pt,bburx=554pt,bbury=762pt}
\caption{
Model $^{13}$CO$~1\rightarrow 0$ line profiles for a Black-Draine
radiation field modified such that the CMB temperature is that
appropriate for high redshift. From top left to right the redshifts are: 0, 1, 2, 3}
\label{cmb3}
\end{figure}
\begin{figure}
\psfig{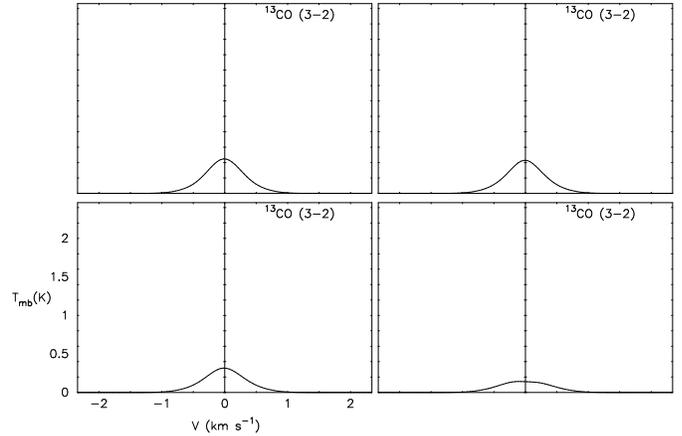}
\caption{
Model $^{13}$CO$~3\rightarrow 2$ line profiles for a Black-Draine
radiation field modified such that the CMB temperature is that
appropriate for high redshift. From top left to right the redshifts are: 0, 1, 2, 3}
\label{cmblast}
\end{figure}
Clearly the gas in a globule of thermal temperature of around 10K
embedded in a medium with a thermal temperature of say, 20K, will
absorb the background radiation via molecular lines. What has been
overlooked before however is that even a low temperature background
will affect the line profiles. Since the separation of energy levels
in many of these species is of the order of a few Kelvin, extra
external photons will have the effect of increased pumping of the
energy levels of the molecules. This results in increased absorption
in the lower J levels and emission from high J levels.

\section{Implications and observational tests}
The results presented here suggest that in most cases interpretations
of fits to observational line profiles will be satisfactory since the
background radiation field does approximate the CMB. However, for an
isolated Bok globule in the vicinity of some newly formed massive
stars, for example, characterising the incident as well as the local
radiation field will be an indispensable first step in modelling the
line profiles. Our adoption of a grey body isotropic addition to the
radiation field is clearly a simplification. It seems likely that the
use of non-isotropic background radiation field could yield line
profile shapes that vary in strength across a source.

Several straight forward observational tests could be carried out to
verify the effects described here. Firstly, a sample of protostellar
cores of similar evolutionary status but situated in quite different
ambient medium conditions could be observed. The prediction is that
the objects in the hotter environments that lead to significant dust
heating and re-radiation at millimetre and sub-millimetre wavelengths
will have weaker molecular emission in the low-lying transitions as
compared to their more embedded counterparts. However, the strength of
the higher level transitions (such as HCO$^+$J=$4\to3$) will be
marginally less affected.  Observations of several lines could
therefore be used to diagnose the environment of star-forming
regions. A second test would be to identify a source that is
non-uniformly illuminated by the background radiation e.g. an object
lying close to the edge of a massive star forming region.  The
morphological distribution, as traced by the sort of low-lying
transitions described above will be biased by the effects described in
this paper and should result in the spatial peak of the molecular
distribution being offset (towards the cooler side of the object) from
the density (continuum) peak. It is possible that the effects
discussed in this paper may be partly responsible for some of the
offsets between the spatial peaks in the various molecular tracers and
the continuum that have been observed towards certain star-forming
regions. It should be cautioned however that it will be difficult to
separate these effects from those of variations in chemical abundances
and kinetic temperature.

One example of a star-forming location in which the external radiation
is likely to differ significantly from the CMB is the Eagle
nebula. \citet{white.et.al99} detected sub-mm emission peaks towards
the heads of the famous columns. They found that the dust temperature
ranges from 250-320K in the hot outer layers to 10-20K deep in the
columns. Clearly, the evaporating gaseous globules
\citep{hester.et.al96} on the outside of the columns will be subject to
much harsher environment than to clumps in the sub-mm column and to
isolated B335-like cores. Similarly, \citet{beuther.et.al00} find a
temperature gradient from 70K to below 20K across the Cepheus B
molecular cloud which, like the Eagle, has a massive star heating one
side of the cloud.

Most interestingly, \citet{white.et.al99} note with concern that in
the sub-mm peaks in the columns of the Eagle nebula, the (1-0)
transitions in CO and its isotopes are all 2-3 times weaker than would
be expected based on the strengths of higher transitions. This has
implications for the common practice of using CO and $^{13}$CO to
measure cloud masses. White et al suggest that freeze-out of CO may
have occurred in the coldest parts of the core. Our results may also
naturally explain these anomalies - the lowest transition is being
weakened by changes to the level populations due the radiation field
from the nearby massive stars.

The higher CMB temperature at high redshift has a clear effect on the
line profile shapes and the suppression of the line strengths could
potentially lead to extra difficulties in the detection of molecular
clouds at redshifts $z>1$. 

\section{Conclusions and future work}
It has been argued that the intensity and shape of submillimetre
molecular line profiles as modelled by radiative transfer codes of
prestellar and protostellar cores are sensitive to the long-wavelength
ambient radiation field illuminating the exterior of the cloud. Some
caution is therefore suggested in the modelling of observational
results - the ambient environment must somehow be constrained before
fitting profiles and making conclusions about the dynamics and
chemical composition of the system.

It should be noted that the effect described here is not unknown to
workers in computational radiative transfer (it was discussed at the
benchmarking exercise that led to the paper by Van Zadelhoff et al
2002), in preparation). However, this is the first time that the
astrophysical consequences of this effect have been explored in any
detail. In future papers on star-formation, we will - following the
preliminary study of \citealt{ward-thompson&buckley01}) - address the
issue of how the velocity structure in general, and the microturbulent
velocity in particular, affects the line profiles. It is important to
realise that it is essential to characterise correctly the
microphysics of potential infall sources. This will eventually allow
more robust inferences to be made from line profile data than is
currently possible.

\section*{Acknowledgements}
MPR and DAW were supported by PPARC and the Leverhulme Trust
respectively while this work was carried out. Some of the calculations
described here were carried out on the Miracle Supercomputer, at the
HiPerSPACE computing Centre, UCL, which is funded by the UK Particle
Physics and Astronomy Research Council.


\label{lastpage} 
\end{document}